Metrics to quantify surrogacy for survival in metastatic oncology trials: a simulation study

Wei Zou

Genentech, S. San Francisco, California, USA

Email: zouw2@gene.com

Funding: The work is supported by my job as a statistician for Genentech. But the company is not directly involved in the design and analysis of the simulation study, the interpretation of the results or in writing the manuscript.

Acknowledgements: The author thanks Jane Fridlyand, Ching-wei Chang, Jue Wang, Rene Bruno and Francois Mercier for methodological and statistical advice.

## Abstract

Surrogate endpoint (SE) for overall survival in cancer patients is essential to improving the efficiency of oncology drug development. In practice, we may discover a new patient level association with survival, based on one or more clinical or biological features, in a discovery cohort; and then measure the trial level association across studies in a meta-analysis to validate the SE. To understand how well various patient level metrics would indicate the eventual trial level association, we considered causal biological trajectories based on bi-exponential functions, modeled the strength of their impact on survival hazards via a parameter $\alpha$, and simulated the trajectories and survival times in randomized trials simultaneously. We set an early time point in the trials when the trajectory measurement became the SE value. From simulated discovery cohorts, we compared patient level metrics including C index, integrated brier score, and log hazard ratio between SE values and survival times. We assembled multiple simulated studies to enable meta-analyses to estimate the trial level association. Across all the simulation scenarios considered here, we found tight correlations among the three patient level metrics and similar correlations between any of them and the trial level metric. Despite the continual increase in $\alpha$, both patient and trial level metrics often plateaued together; their association always decreased quickly as $\alpha$ increased. This suggests that incorporating additional biological factors into a composite SE is likely to have diminishing returns on improving both patient level and trial level association.

## Introduction

Cancer immunotherapies (CIT) have significantly extended the overall survival (OS) for oncology patients and become the new standard of care in many metastatic settings. This means future drug development in these settings will take even longer to read out OS benefit; and hence a stronger need for better surrogate endpoints (SE) for OS. SE is supposed to be cheaper/easier/quicker to measure [1] than OS itself. If the treatment benefit measured on SE can reliably 'predict' the OS benefit from the treatment, we can then infer OS benefit in a cheaper/easier/quicker way. We can quantify the reliability using an $R^2$ from a meta-analysis of such pairs of treatment effects across many clinical trials [2,3]. Endpoints based on Response Evaluation Criteria in Solid Tumors (RECIST), e.g., objective response rate and progression-free survival have various levels of $R^2$ with OS in different settings but are rarely considered validated SE [4,5]. They have nevertheless led to accelerated approval in specific contexts [6]; they are also indispensable in go/no-go decisions in early clinical development in oncology. Further improving SE for OS is critical to bring effective novel treatments to cancer patients more efficiently.

OS for cancer patients is affected by many prognostic factors [7]. Patients' immune system in general and in the tumor microenvironment play important roles to derive OS benefit under CIT, besides cellular property of tumors [8]. The complex biology strongly advocates that we may improve SE by starting from a feature pool covering comprehensive biology of cancer, patient and treatment, and screening for feature combinations that can predict OS risks using statistical learning / predictive modeling methods [9]. Following the SE validation framework from Buyse et al [2,3], we will first train/discover a patient level OS model in one large cohort with comprehensive data collection. This step can use all the feature selection or prognostic modeling approaches under a landmark analysis framework [10] to model the relationship between OS and baseline or early on-treatment variables. Then we can validate the utility of a novel feature or model as a SE in a group of studies where the necessary variables are consistently measured. $R^2$ from the meta-analysis measures the utility of a SE [2].

Unlike $R^2$ as the clear choice to measure a trial level association, there are several different metrics to describe a patient level association with OS: C index, Integrated Brier score (IBS), and hazard ratio of the SE in a Cox proportional hazard model. C index measures the discrimination performance, i.e., the probability that patients with larger SE values have shorter survival times than those with smaller values [11]. It can be interpreted as a weighted average of the area under curve (AUC) of a time-dependent Receiver operating characteristic (ROC) curve [12]. In contrast, IBS is a weighted average of time-dependent brier score (BS), where BS is the square error between predicted survival probabilities and actual survival status at a time. These two metrics are commonly used to measure the performance of complex prediction models for time-to-event endpoints [13,14]. Besides, many SE discovery applications reported Kaplan Meier (KM) curves or hazard ratio (HR) of the SE to illustrate the prognostic values [5,15,16].

To understand the relationship among the metrics to measure OS model performance at the patient level as well as their relationship with the trial level $R^2$ in SE development, here we simulated the path from patient level associations to trial level associations and studied their association. This work follows the spirits of previous works [17,18] to interpret the values of patient level metrics and their improvement using the scale of the eventual utility of a SE.

## Methods

**Data simulation**

We used a joint model (JM) framework [19,20] to simultaneously model the underlying biological trajectory $f_i(t)$ for patient $i = 1..N$ at time $t$ and the OS event hazard as a function of the status:

$$Y_i(t) = f_i(t) + \varepsilon_i$$

$$h_i(t) = \gamma t^{\gamma-1} \exp\{\beta_0 + \beta_1 trt_i + \alpha f_i(t)\}$$

In the longitudinal submodel, $f_i(t)$ encodes the pharmacodynamics (PD) modulation on the underlying status from the treatment (more details are in the next paragraph); $\varepsilon_i$ represents the measurement error that does not impact the hazard and $Y_i(t)$ is the apparent value of $f_i(t)$. The longitudinal submodel is linked to the Weibull time-dependent hazard submodel via a coefficient $\alpha$, which measures the strength of the biological link between the two. We conducted simulations where $\alpha$ was 0, 0.5, 2, 4, or 6. Although $\alpha$ has to be formally estimated from a JM fit, the log HR from a univariate Cox model between OS and $Y_i(t)$ may provide some hint of its magnitude. We may often be able to identify one prognostic factor with $\alpha$ around 0.5 in a treatment setting (e.g., ctDNA features in 2nd line CIT naïve NSCLC [21]). If we are lucky, we may gather multiple independent factors, each with $\alpha$ around 0.5, and get close to $\alpha = 2$ after combining them. $\alpha = 6$ in log scale is a huge effect. As the treatment, $trt_i$ for patient $i$, may affect hazards by altering additional biological pathways not captured by $f_i(t)$, we used $\beta_1$ to code such a 'leaked' treatment effect, with its value being 0, -0.3, or -0.6 in the simulations. When $\alpha$ deviates from 0 and $\beta_1$ moves towards 0, $f_i(t)$ captures more and more biological variability underlying OS hazards. Such changes could mimic a spectrum of multi-factor models that gradually pull relevant features in to predict OS. Additional parameters in the hazard submodel include $\gamma$, which is set to 1 unless specified otherwise; $\beta_0$ the baseline hazard; and $trt_i = 1$ for patients in an active treatment arm and 0 for those in a control arm.

We used bi-exponential functions as $f_i(t)$ as they are flexible to describe various forms of PD modulation in the metastatic settings. Since Stein *et al* applied such a function to model the tumor growth inhibition (TGI) in renal cell carcinoma[22], the Stein model has been applied to the sum of longest diameters of target lesions (SLD) in non-small cell lung cancer [23], organ-specific tumor lesion sizes in colorectal cancer [24], or prostate specific antigen in metastatic castration-resistant prostate cancer [25].

$$f(t) = \exp(-K_s t) + \exp(K_g t) - 2$$

There is a Ks parameter to describe the initial drop (or tumor regression in the Stein model), reflecting temporary treatment benefit; and a Kg parameter for later increase (or tumor growth), capturing the eventual treatment failure common in the metastatic settings. Unlike the Stein model which describes the observed lesion sizes, $f_i(t)$ models the proportion of change from baseline, and is relevant to a broad category of biological processes after proper transformation. We simulated individual Ks/Kg values (unit week$^{-1}$) from a lognormal distribution with a variance of 0.8 for Ks and 0.6 for Kg, plus a measurement error term $\varepsilon_i \sim N(0, 0.09^2)$. These values were from the TGI model estimates in POPLAR, a randomized trial in metastatic NSCLC [23]. So the PD modulation in the simulations had similar random patient effects as the TGI fit to POPLAR. To simulate a range of fixed treatment effects on $f(t)$, we covered a grid of means for Ks/Kg from 0.01-0.02 (from TGI models in POPLAR) to more extreme values (Table 1) to accommodate SE with a similar or larger dynamic range as compared to SLD / RECIST assessments, or treatments with similar or stronger PD effects. We anchored the simulation parameters on the TGI parameters for SLD because they are an obvious baseline to develop new complex SE for OS in metastatic settings where tumor response has been playing an important role in patient management, drug development, and regulatory approval.

In simulation scenarios Ks1 or Kg1, control arms had similar trajectories as the TGI model from POPLAR; while the active arm could be superior in Ks or Kg, respectively. Scenario Ks1-low and Ks1-high were similar to Ks1, except that Ks1-low had the mean Ks in the active arm varying between 0.02 and 0.04; and Ks1-high between 0.04 and 0.06. Simulation Ks2 was another sensitivity analysis to Ks1 by increasing the common Kg rates across both treatment arms. Simulation Kg2 was a sensitivity analysis to Kg1 by increasing the common Ks across arms.

Given $f_i(t)$ and hence $h_i(t)$, we simulated OS times by finding event times so that their cumulative distribution is a uniform one via the numerical solution from the R package simsurv [26]. Unless specified otherwise, each study randomized 200 patients to each of the two treatment arms, to mimic a phase III study with a target OS HR of 0.72. We adjusted $\beta_0$ in different simulations so that there were not too many early events (< 25% of patients) before the time point to measure the SE values (discussed later) and yet the overall trial duration was realistic (< 120 months with a final event/patient ratio of 0.75). We did not apply additional censoring mechanisms. We simulated 300 studies under each parameter combination.

To estimate the trial level association, we randomly sampled a group of studies to simulate a meta-analysis dataset, each including all possible values of Ks/Kg under one scenario in Table 1. It is necessary to include studies with heterogeneous PD modulations to estimate the trial level association [2]. We then paired the group, as well as the estimated trial level association, with a random SE discover study (not included in the meta-analysis) where we estimated the patient level association. We stratified the sampling and pairing step by both $\alpha$ and $\beta_1$ (i.e., all studies contributing to a pair had the same value of $\alpha$ and $\beta_1$), or just by $\alpha$ (i.e., $\beta_1$ took all values from 0, -0.3, or -0.6 in the meta-analysis and took a random value from 0, -0.3 or -0.6 in the SE discover study). Within each scenario, we simulated 100 pairs under each unique combination of $\alpha$ and $\beta_1$ values, or just each unique $\alpha$.

## Data analysis

$f_i(t)$ varies across time. In practice, each SE needs a particular time point $t^*$ not too late into the study when $Y_i(t = t^*)$ is the SE value for patient $i$. Because the parameter values for $f_i(t)$ here were originated from a TGI fit on SLD data, we set $t^* = 2$ months, which is roughly the time of the 1st on-treatment RECIST assessment [27] as well as many other measurements in metastatic trials, e.g., circulation tumor DNA [28] or C reactive protein [29]. We envision simulations here are directly relevant to SE development in many metastatic settings where a novel SE at around 2 months will be benchmarked against and/or combined with the 1st on-treatment radiographic assessment in predicting OS.

In each simulated study, we estimated the patient level association between $Y_i(t = 2)$ and OS time using Harrell's C index [11] across treatment, Integrated Brier Score (IBS) [13] across treatment, or just the log HR of the SE in a Cox model stratified by treatment. Values of C index and IBS by treatment are highly correlated with the statistics summarized across treatment hence not reported here. In IBS calculation, survival probabilities were estimated from a Cox model with $Y_i(t = 2)$ value as the only predictor; censoring probabilities were from a Cox model with intercept only. A scaled IBS [30] normalizes an IBS against its value under an intercept-only prediction model to correct for the amount of censoring. In assessing the patient level association, we excluded patients with OS events before $t^*$ and re-baseline OS to $t^*$ to avoid immortal bias[15,31].

To estimate the trial level association, we estimated the marginal treatment effect on OS as the log HR of $trt_i$ in a univariate Cox model, and the treatment effect on the SE as the difference of median $Y_i(t = 2)$ between treatment arms in the full population of each study; and then estimated the coefficient of determination (i.e., $R^2$) between the 2 estimates across the meta-analysis dataset. For patients with a simulated OS event before $t^*$, we set their SE values as the maximum SE value in the at-risk population.

## Results

### Patient level associations

Figure 1 shows C index and IBS values under scenario Ks1. We first considered correlations across studies simulated under the same parameter values. In Figure 1, such studies are in the same small panel and have the same color. There appear strong correlations between the two metrics when $\alpha$ is above 0.5. Similar patterns are observed for log HR of SE. The correlations are especially tight between the C index and scaled IBS, which are generally above 0.9 as long as $\alpha$ is above 0 in all scenarios evaluated. In supplement file 2-7, Section 'correlation summarized from studies simulated under identical parameters' includes similar visualization as Figure 1 for other metrics. Figure S1 in supplement file 1 summarizes such correlations across scenarios.

As we may use these patient level metrics to rank the strength of the biological link between a SE and OS, the correlations obtained when $\alpha$ varies (i.e., clouds of different colors within a small panel in Figure 1) are more meaningful. The median absolute values of such spearman correlations between the C index and other metrics are usually 0.95 or larger in all scenarios (Figure 2). The tight correlations appear consistent across different values of $\beta_1$ and Ks/Kg, given the narrow interquartile ranges in Figure 2.

Though $\alpha$ drives the C index as expected, C index values start to plateau around 0.7 ~ 0.8 once $\alpha$ reaches 2 (Figure 3). Given $\alpha$, different Ks/Kg appear to have a minor impact, judging from the interquartile ranges; $\beta_1$, the leaked treatment effect, has little impact on the patient level association, especially when $\alpha$ is large. A non-zero $\beta_1$ may slightly inflate the C index when $\alpha$ is close to zero. There are similar plateaus between $\alpha$ and IBS or log HR of SE (figure S4 – S6 in supplement file 1).

**Trial level associations**

**Fixed $\beta_1$ in meta-analyses**

Under an ideal setting where we have 15 studies to estimate the trial level association, where all 15 have the same underlying $\alpha$ and $\beta_1$, and replicate each of the 5 different Ks/Kg values three times (See examples of treatment effect estimates to support meta-analyses for scenario Ks1 in Figure 4), $\alpha$ drives such trial level association as expected as well (Figure 5). In all scenarios, however, $R^2$ plateaus, often well below 1, despite the continual increase in $\alpha$; the plateau consistently occurs when $\alpha \geq 2$, just like those patient level metrics (e.g., *Figure 3*). $R^2$ tends to decrease when $\beta_1$ deviates from zero, though the variation within a $\beta_1$ value is much larger. When we reduce the range of Ks in the meta-analysis in Scenario Ks1 (hence reducing the dynamic range of treatment effects on the SE), $R^2$ becomes lower in Scenario Ks1-low and particularly lower in Ks1-high.

Figure 6 suggests there are similar correlations between any of the patient level metrics and the trial level $R^2$ in all scenarios; studies where $\alpha \leq 0.5$ have a major contribution to the correlations. In scenario Ks1, for example, the correlations decrease from 0.8 (considering studies with $\alpha$ varying from 0 to 6) to around 0.2 (considering only those with $\alpha \geq 2$). Section 'Fixed beta_1' in supplement information file 2-7 includes scatter / contour plots between the patient and trial level metrics behind these correlations. The leaked treatment effect $\beta_1$ tends to reduce their correlation slightly. When we reduce the range of Ks in the meta-analysis in Scenario Ks1, correlations in Scenario Ks1-high become much weaker, echoing the much-reduced $R^2$ in this scenario (Figure 5); yet the correlations barely shift in Scenario Ks1-low.

In a less ideal setting where we only have 5 studies to estimate the trial level association, where each of the 5 different Ks/Kg values in a scenario just has 1 replicate, there is more variability in $R^2$ (figure S9 in supplement file 1). For example, the median $R^2$ becomes 0.17 when both $\alpha$ and $\beta_1$ are zero under simulation Ks1 here, while the corresponding 15-study $R^2$ values are more squeezed to zero (Figure 5) as they should be. The less noisy 15-study $R^2$ could illustrate the true patterns more clearly. Although there is an overall decrease in correlation between the patient and the trial level association with 5 studies in the meta-analysis, the $R^2$ plateau and correlation decrease remain the same when $\alpha \geq 2$ (figure S10 in supplement file 1).

**Varying $\beta_1$ in meta-analyses**

It is possible that studies to estimate a trial level association have different $\beta_1$ values but still the same $\alpha$ as the SE discovery cohort. Section 'Varying beta_1' in supplement file 2 -7 includes examples of treatment effect estimates to support meta-analyses in each scenario. With this setup, there is no consistent plateau in $R^2$ when $\alpha$ increases (Figure 7). However, $R^2$ remains similarly correlated with all the patient level metrics; and excluding simulations with $\alpha \leq 0.5$ still remarkably reduces the correlation between the patient and trial level metrics (Figure 8).

Discussion

In this paper, we simulated metrics for the patient level association and paired them with the trial level $R^2$ to understand the differences among various patient level metrics and the significance of their numerical improvement on the eventual utility as SE under known data generation mechanisms. From the simulation, it is striking that the following patterns remain clear across the scenarios: all 3 patient level metrics are highly correlated when $\alpha$ varies, and the following happens after $\alpha$ reaches a certain value:

- all the patient level metrics plateau similarly;
- the trial level $R^2$ often plateau as well;
- there is little association between these two categories of metrics.

**Simulation setup**

In real practice, at the time of discovering the patient level association using statistical learning methods, we would not know what studies are included to assess the trial level association: how many studies would be available, what the range of PD modulation would be, or whether $\beta_1$, the leaked treatment effect, would remain the same as that in the discovery cohort. Our simulation suggested all of these could affect the absolute value of $R^2$ and its association with a patient level metric. Despite all the uncertainties/heterogeneities, we still need $R^2$ from separate meta-analyses to interpret the numeric value of any patient level association metric, or its improvement when we update the predictive model underlying the SE (e.g., including additional prognostic factors) in the initial discovery cohort. This makes analytical modeling very hard and the parametric simulation an appropriate and pragmatic approach to start addressing this problem.

We tried to characterize the relationships among the metrics in ideal settings. We assumed we knew when and how to measure the SE, with certain measurement errors though. We assumed the biological link between the SE and survival in a treatment setting was generalizable across studies so that we kept $\alpha$ the same when we randomly paired the initial discovery cohort and the meta-analysis dataset. We allowed $\beta_1$ to be the same or different. We used a range of Ks/Kg parameters to cover heterogeneities among treatment options, and perturbed one parameter at a time across different scenarios (Table 1). Simulation results from these ideal settings, however, are directly relevant to the statistical learning practice in certain SE development contexts, especially to metastatic NSCLC. The simulation approach remains relevant after adjusting parameter values to fit alternative contexts. All simulation programs are available at https://github.com/zouw2/oc_metrics.

**Patient level association metrics for survival are tightly correlated**

Our simulation confirmed the correspondence between the C index and the log HR of SE. The D statistic by Royston et al [32] can be interpreted as log HR and Jinks et al [33] derived an empirical function between C index and D statistics based on literature review. Our simulation also demonstrated a tight association between the C index and IBS (or its scaled version) under all scenarios considered. More importantly, they are similarly correlated with $R^2$. This appears a surprise given that the C index only measures rank concordance or discrimination; while IBS measures both calibration and discrimination performance [34] in the patient level association.

One contributing factor to the lack of distinction could be that the trial level association is based on the relative treatment benefit estimated from randomized studies in general, and on HR from a Cox model in oncology drug development. The setting to find SE is different from a typical predictive modeling application for a binary endpoint, where predicted absolute risks directly affect the utility [17,18]. Additionally, since Cox likelihood is the same as the marginal likelihood of ranks of observed times (when there is no censor), a SE with a perfect rank concordance with survival in a randomized trial could contain the full information about the marginal treatment effect on survival.

**SE Association is hard to improve by recruiting more causal factors**

The simulation recovered one difficulty which we frequently came across in practice: the C index for survival is hard to improve by bringing new predictors into an existing model, just like its equivalent AUC for a binary outcome [35,36]. This suggests that we will likely have the same difficulty in the ideal settings simulated in this work. Here we increased $\alpha$ to mimic ideal multivariate modeling applications where we were able to incorporate additional casual variables for survival into the SE. Still, the C index grew very slowly after around 0.7 while $\alpha$ increased from 2 to 6 (Figure 3). Other patient level metrics plateaued similarly. In many simulation settings (all scenarios when $\beta_1$ was fixed in meta-analyses in Figure 5, or Scenario Kg1 and Kg2 when $\beta_1$ varied in meta-analyses in Figure 7), trial level $R^2$ had the plateau coincidently with the patient level metrics. In all scenarios, the association between the patient level and trial level metrics became much weaker after $\alpha \geq 2$ (Figure 6 and Figure 8), suggesting that further improving the patient level association may have little utility. The lack of association was sometimes linked to reduced $R^2$ (e.g., Scenario Ks1-high) but seemed always influenced by the plateau in the patient level metrics. For example, although Scenario Ks1 or Ks2 had a nice linear trend between $R^2$ and $\alpha$ when $\beta_1$ varied in meta-analyses (Figure 7), the plateau in patient level metrics (Figure 3) suggested the continual increase in $R^2$ after $\alpha \geq 2$ in these scenarios was not driven by the stronger patient level association in the discovery cohort.

There could be a common bottleneck for both the patient and trial level association in the simulation. One suspect is the limited information we can gather at the early time point $t^*$ when the SE is measured. A stronger $\alpha$ not only strengthens the biological link between $f_i(t)$ and OS hazards from time 0 to $t^*$; but also amplifies the impact of the future variations in $f_i(t)$ that are not captured at $t^*$. Our $f_i(t)$ often had a U shape under the bi-exponential function (Section 'trajectories' in supplement file 2-7), which reflected the non-curable nature of metastatic diseases and suggested variations after $t^*$ could be substantial. This might explain the clearer $R^2$ plateau (Figure 5) in Scenario Kg1/Kg2 compared to other scenarios where the Kg (i.e., the growth parameter) variation drove different treatment effects. Although defining an early decision time point $t^*$ is a necessary component of SE, further research is needed to understand whether an estimate of the trend of $f_i(t)$ at $t^*$ could be better than observed or change from baseline measurements.

## Conclusions

Under the SE development path considered here, the three patient level metrics C index, IBS and log HR performed similarly to rank candidate SE in terms of their eventual trial level association.

Our simulations caveated the use of multivariate modeling on observed or change from baseline features to improve SE for survival. When α is 2 or above in our simulations, the difficulty to improve patient level metrics like C index may be not an artifact produced by the metric but rather reflect the difficulty to improve the true SE utility; the dramatic reduction in correlation between the patient level and trial level association suggests further improvement in any of the patient level metrics might have a small chance to actually improve $R^2$. Research on alternative modeling strategies to mine oncology trial data to discover SE is warranted.

**Declarations**

- Ethics approval and consent to participate: Not applicable
- Consent for publication: Not applicable
- Availability of data and materials: All simulation programs are available at https://github.com/zouw2/oc_metrics
- Competing interests: Genentech employee and own the company stocks

# Tables

Table 1 Mean Ks and Kg values in different simulation scenarios for the longitudinal submodels

| Scenario | Ks(trt=1) | Ks(trt=0) | Kg(trt=1) | Kg(trt=0) |
|---|---|---|---|---|
| Ks1 | 0.02, 0.03, 0.04, 0.05, 0.06 | 0.02 | 0.01 | 0.01 |
| Ks1-low | 0.02, 0.025, 0.03 0.035, 0.04 | 0.02 | 0.01 | 0.01 |
| Ks1-high | 0.04, 0.045, 0.05, 0.055, 0.06, | 0.02 | 0.01 | 0.01 |
| Ks2 | 0.02, 0.03, 0.04, 0.05, 0.06 | 0.02 | 0.03 | 0.03 |
| Kg1 | | 0.02 | 0.015, 0.012, 0.008, 0.0045, 0.001 | 0.015 |
| Kg2 | | 0.05 | 0.03, 0.025, 0.02, 0.01, 0.005 | 0.03 |

Mean trajectories under each scenario are visualized in the Section 'trajectories' of supplement file 2-7.

Figure Captions

*Figure 1 IBS and C index metrics from Scenario Ks1*

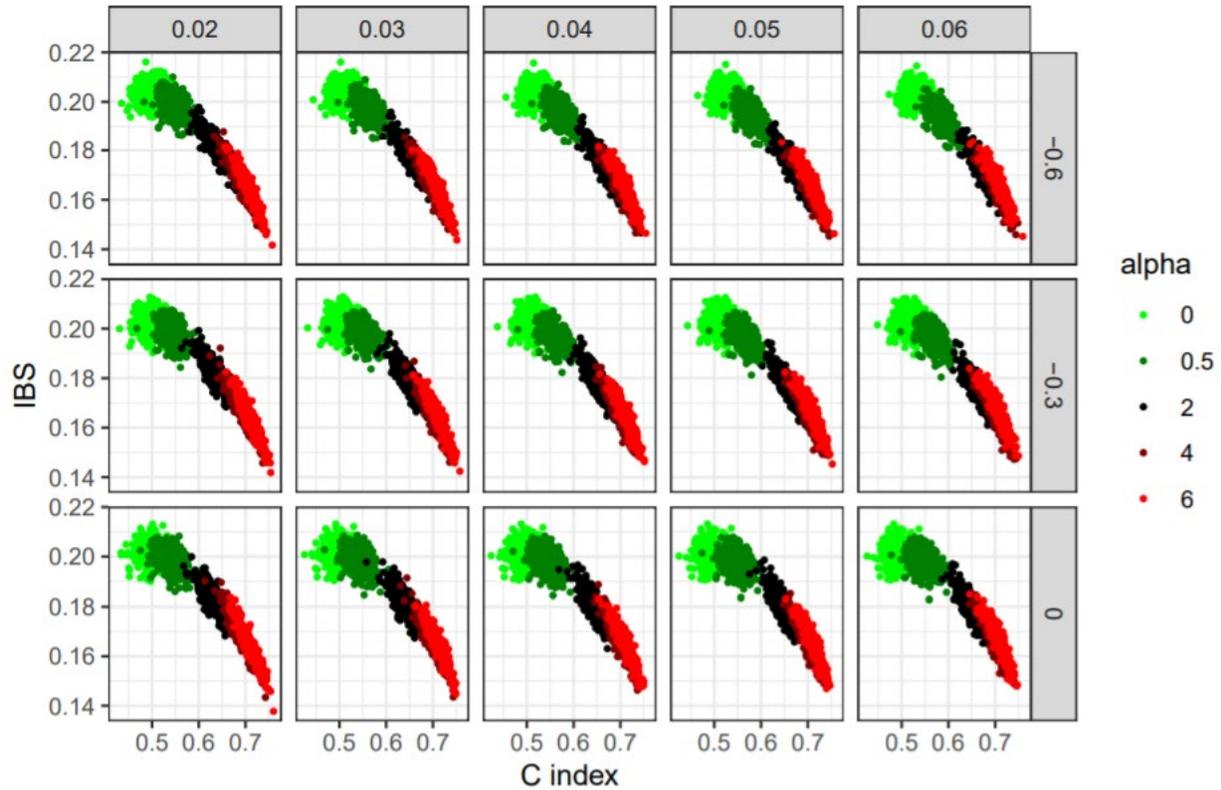

Fig caption: Simulations with different values of Ks in the active arm are arranged along the columns, different values of $\beta_1$ along the rows. Each color indicates one value of $\alpha$. Each cloud of one single color in a small panel includes 300 replicates.

*Figure 2 Correlation between C index and other patient level metrics when alpha varies*

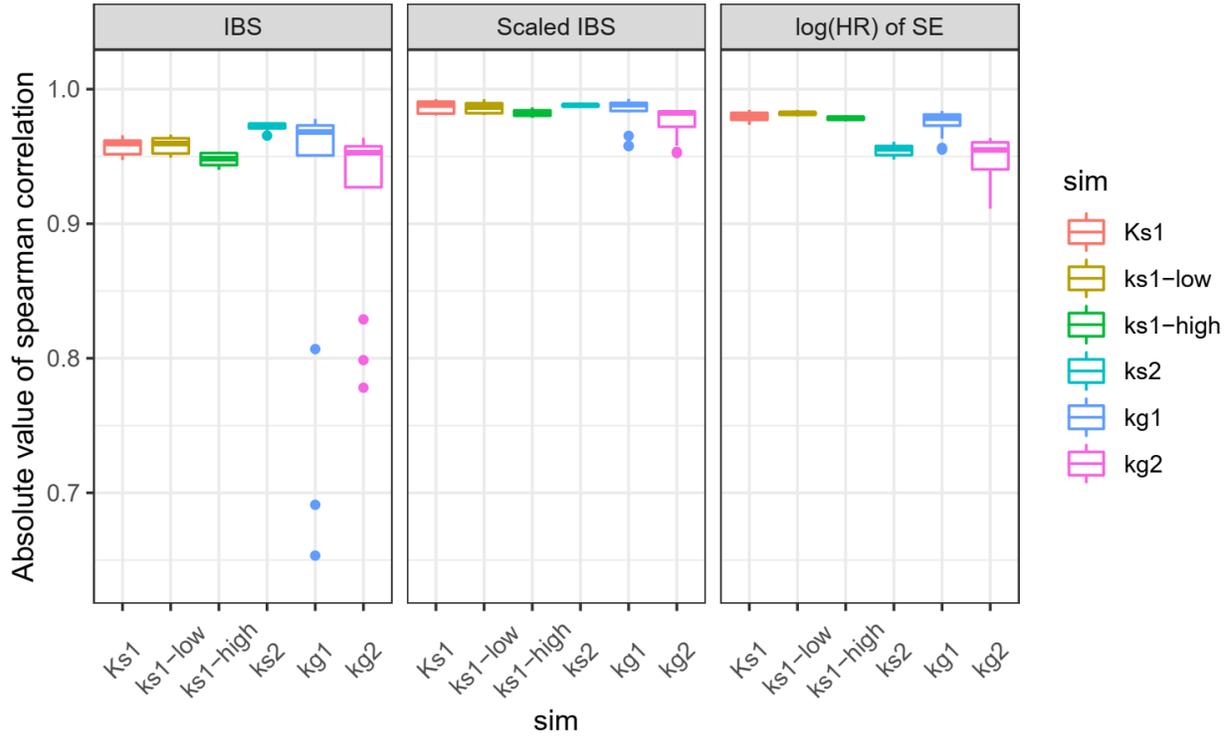

Figure caption: We estimated spearman correlations between C index and other patient level association metrics (indicated in the column header) with $\alpha$ varying and other parameters fixed across 1500 studies (i.e., 300 studies with one of 5 different values of $\alpha$). Each boxplot then summarizes 15 correlation estimates from 3 values of $\beta_1$ and 5 values of Ks/Kg parameter in the active arm under a scenario. Actual correlations are listed in table 3 of the supplement file 2-7 for each scenario.

*Figure 3 C index increases as α, stratified by $β_1$*

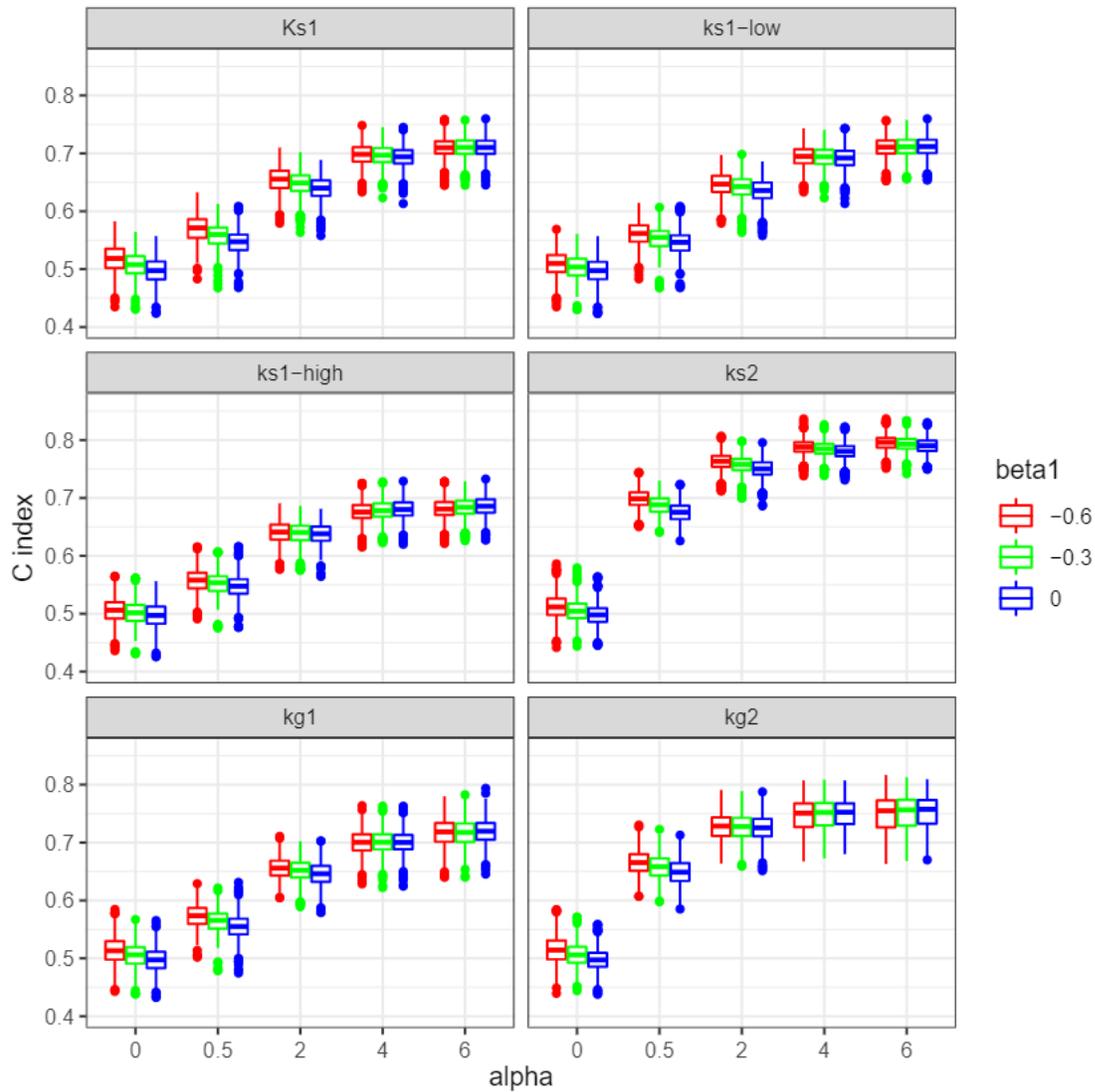

Fig caption: Each boxplot summarizes the C index values from 1500 studies (i.e. 300 studies simulated with one of 5 different values of Ks/Kg parameter for the active arm) given a unique combination of $α$ (indicated on the X-axis) and $β_1$ (indicated by color).

*Figure 4 Examples of meta-analyses to estimate $R^2$, under Scenario Ks1*

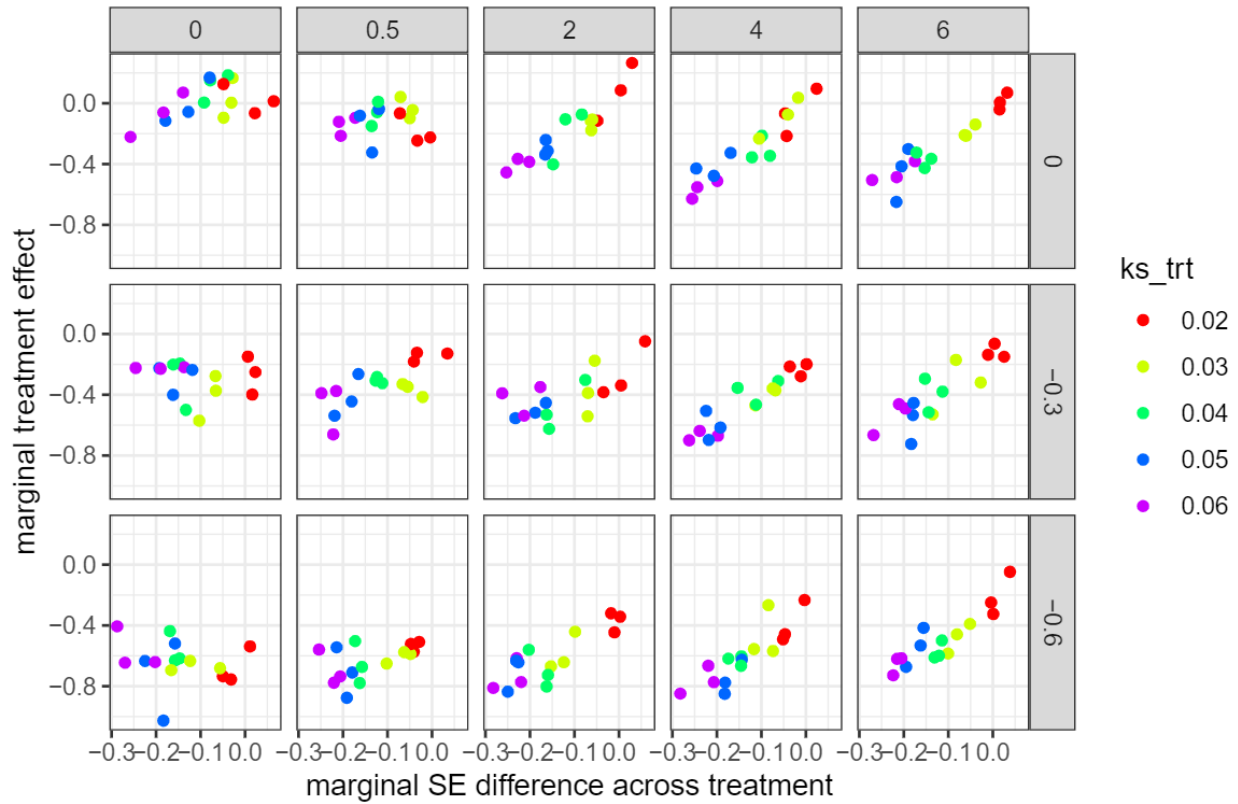

Figure caption: Each small panel illustrates a meta-analysis of treatment effect estimates from 15 studies: each color indicates 3 replicates with the same Ks value in the active arm. $\alpha$ varies by columns and $\beta_1$ varies by rows. The Y-axis shows the log HR for the marginal treatment effect on OS. The X-axis shows the differences in median SE values between the two arms. Each panel will give one $R^2$ estimate.

*Figure 5 $R^2$ distribution by $\alpha$ and $\beta_1$*

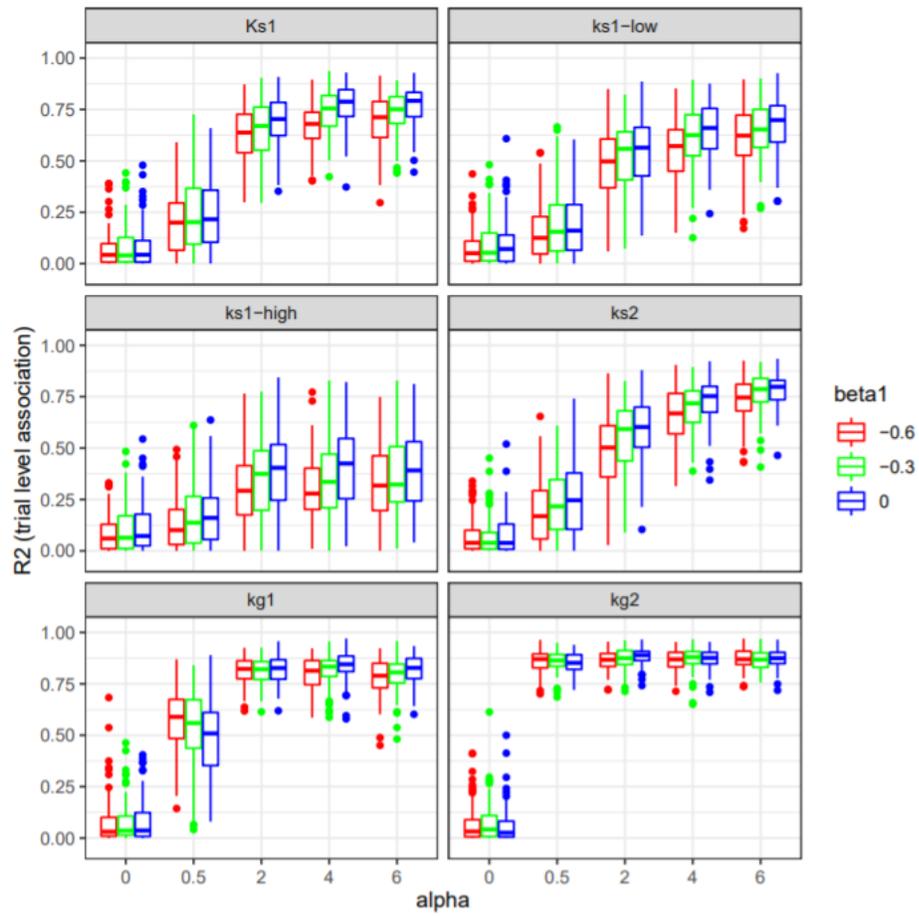

Fig caption: each boxplot summarizes $R^2$ from 100 meta-analyses. Each meta-analysis includes 15 studies, i.e., one small panel in Figure 4.

*Figure 6 Spearman correlation between the patient level and trial level association metrics, stratified by $\beta_1$*

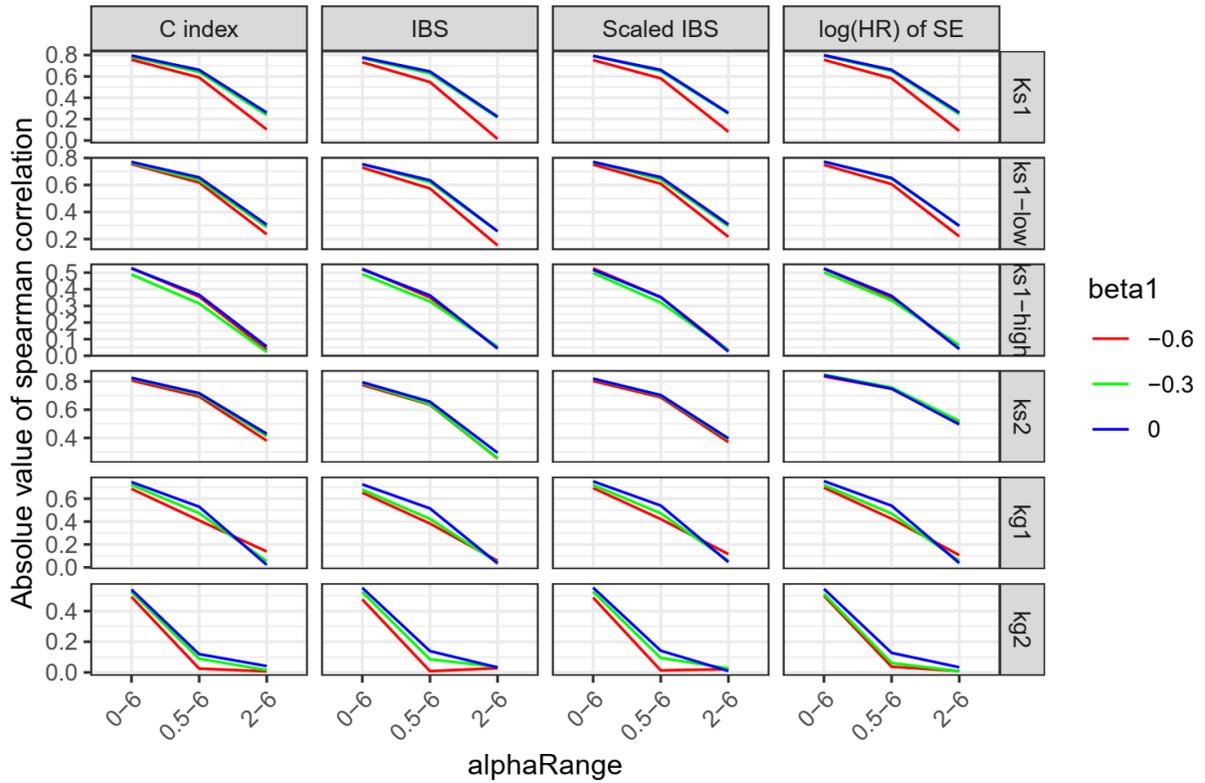

Figure caption: We paired each $R^2$ with a patient level metric (indicated in the column header) simulated under the same of $\alpha$ and $\beta_1$ (indicated by different colors). We summarized spearman correlations across such pairs under **different subsets of** $\alpha$, given a $\beta_1$. We used the full range of $\alpha$ or excluded pairs from studies with $\alpha$ being 0 or excluded pairs with $\alpha$ being 0 or 0.5 (indicated on the X-axis).

*Figure 7 $R^2$ distribution by α, with $β_1$ mixed*

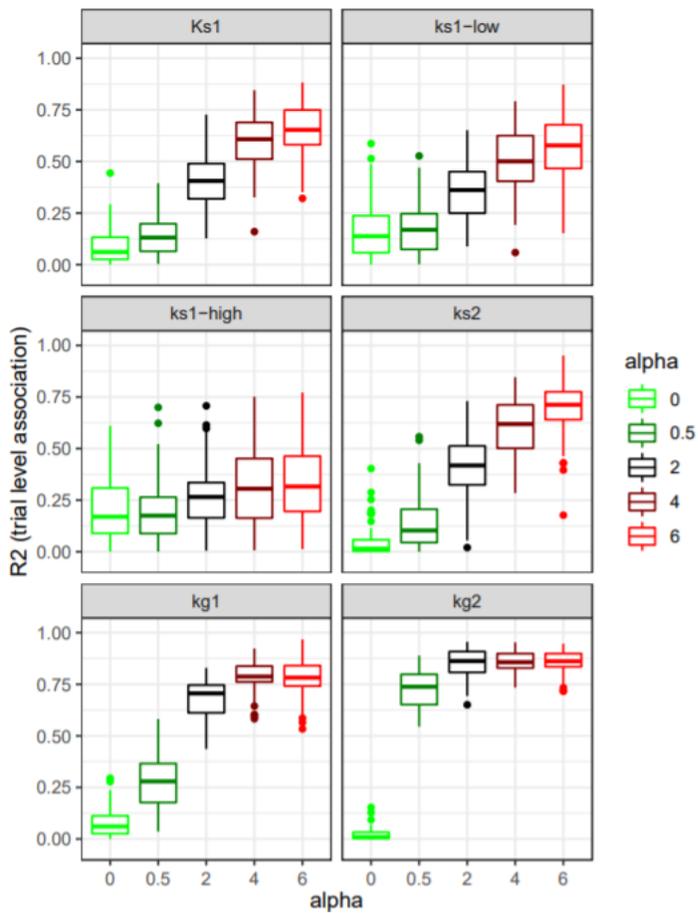

Fig caption: each boxplot summarizes $R^2$ estimated from 100 meta-analyses. Each meta-analysis includes 15 studies, i.e., each simulated under one of three values of $β_1$ and one of five values of Ks/Kg parameter in the active arm.

*Figure 8 Spearman correlation between patient level and trial level metrics, with $\beta_1$ mixed*

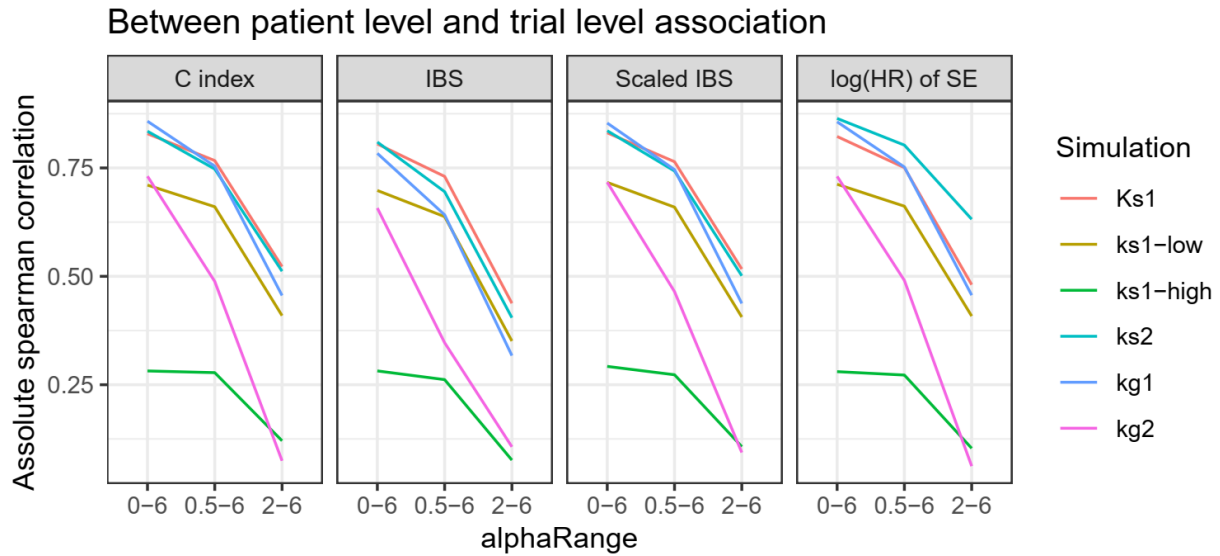

Figure caption: We paired each $R^2$ with a patient level association metric (indicated in the column header) simulated under the same value of $\alpha$ ($\beta_1$ could be different). We summarized spearman correlations across such pairs under different subsets of $\alpha$. We used the full range of $\alpha$ or excluded pairs from studies with $\alpha$ being 0 or excluded pairs with $\alpha$ being 0 or 0.5 (indicated on the X-axis).

## List of supplement files

file_1_sum_across_scens1.pdf: the high resolution version of the figures to be included in the main text or as supplement information

file_2_ks1_output1.pdf: simulation results under scenario ks1

file_3_ks1-low_output1.pdf: simulation results under scenario ks1-low

file_4_ks1-high_output1.pdf: simulation results under scenario ks1-high

file_5_ks2_output1.pdf: simulation results under scenario ks2

file_6_kg1_output1.pdf: simulation results under scenario kg1

file_7_kg2_output1.pdf: simulation results under scenario kg2

# summarize across scenarios

wei zou

2021-05-02 17:52:28

## Patient level association metrics are correlated

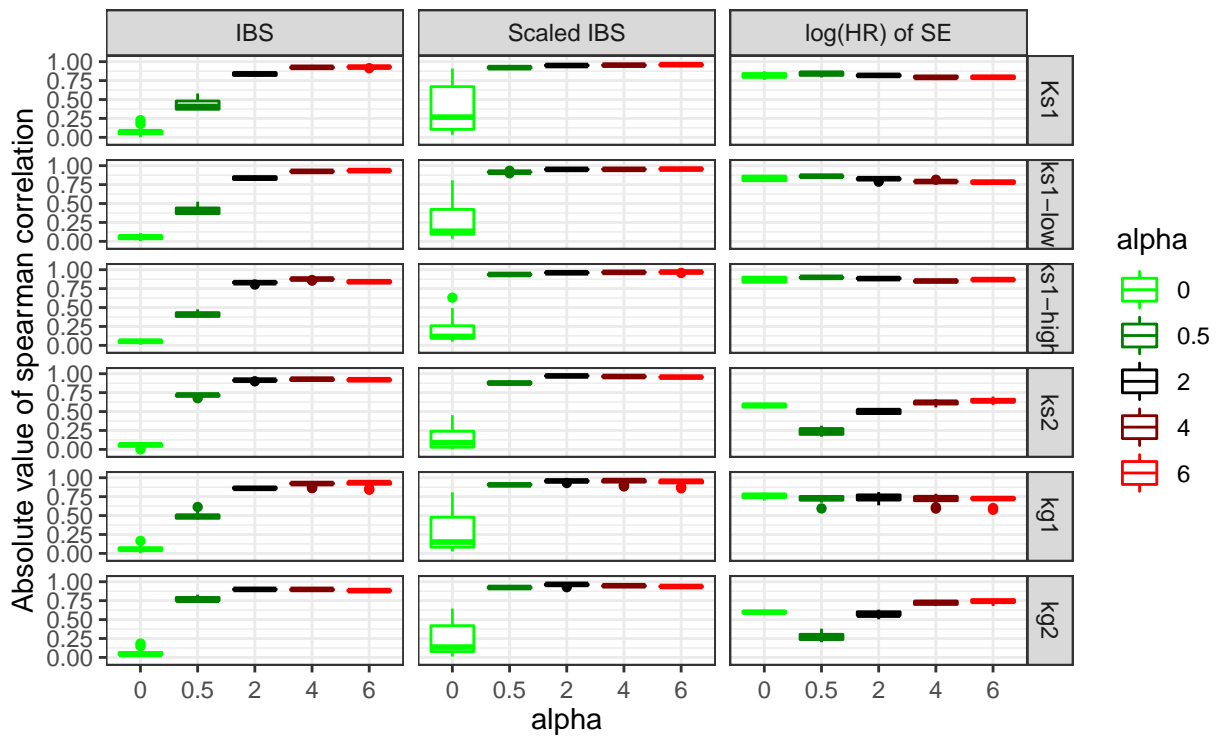

Figure S1: Between C index and other patient level metrics given an alpha

We first estimated spearman correlations between C index and other patient level association metrics (indicated in the column head) from 300 trials simulated from the same parameters. Each box summarizes 15 correlation estimates (from 3 values of beta_1 and 5 values of Ks/Kg parameter in the active arm)

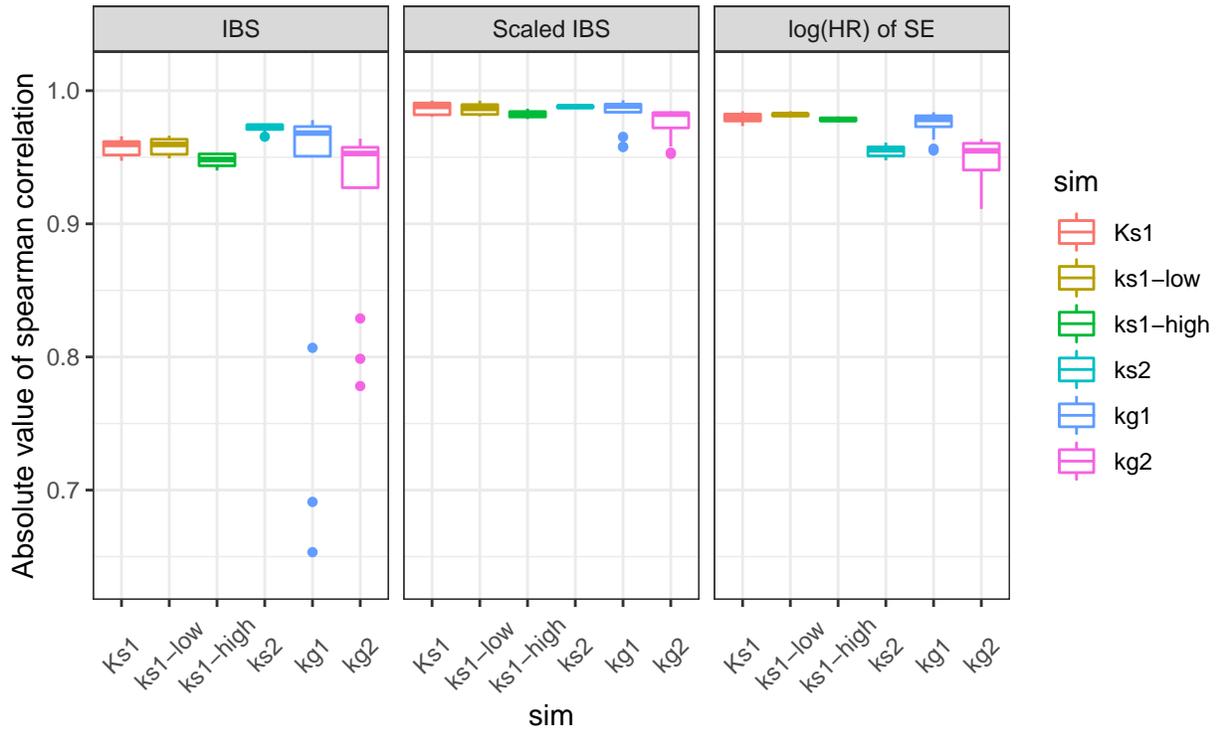

Figure S2: Between C index and other patient level metrics when alpha varies

We first estimated spearman correlations between C index and other patient level association metrics (indicated in the column head) from 1500 studies simulated from the same parameters except that alpha make take 1 of the 5 values in every 300 studies. Each box summarizes 15 correlation estimates (from 3 values of beta_1 and 5 values of Ks/Kg parameter in the active arm)

# patient level association increases with alpha

Figure S3: C index distribution by alpha and beta_1

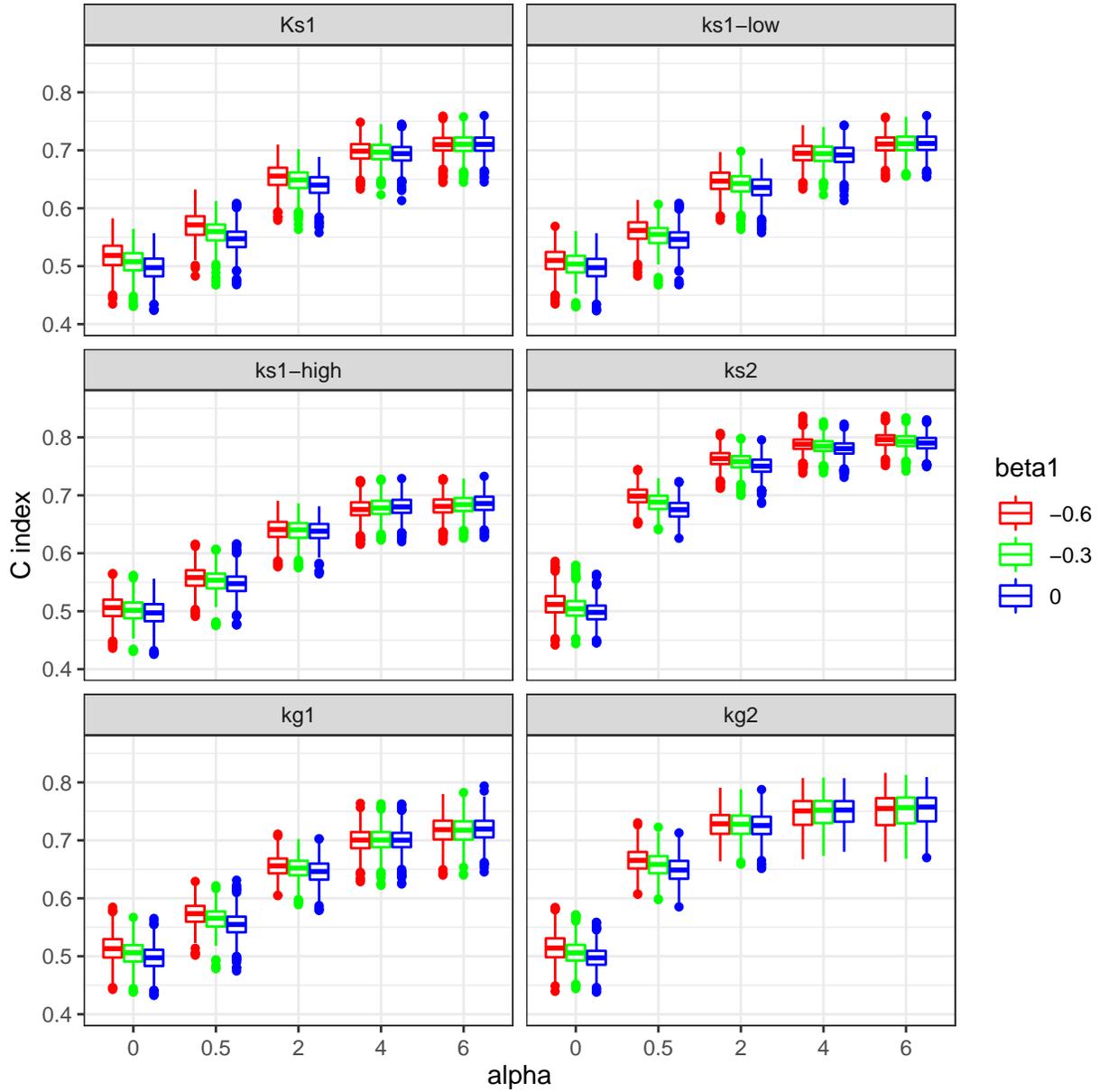

Each box summarizes the 1st quartile, median and 3rd quartile of the C index values from 1500 studies: there are 300 trials simulated with one of 5 different values of Ks/Kg parameter for the active arm and one value of alpha (indicated on the x axis) and one value of beta_1 (indicated by color).

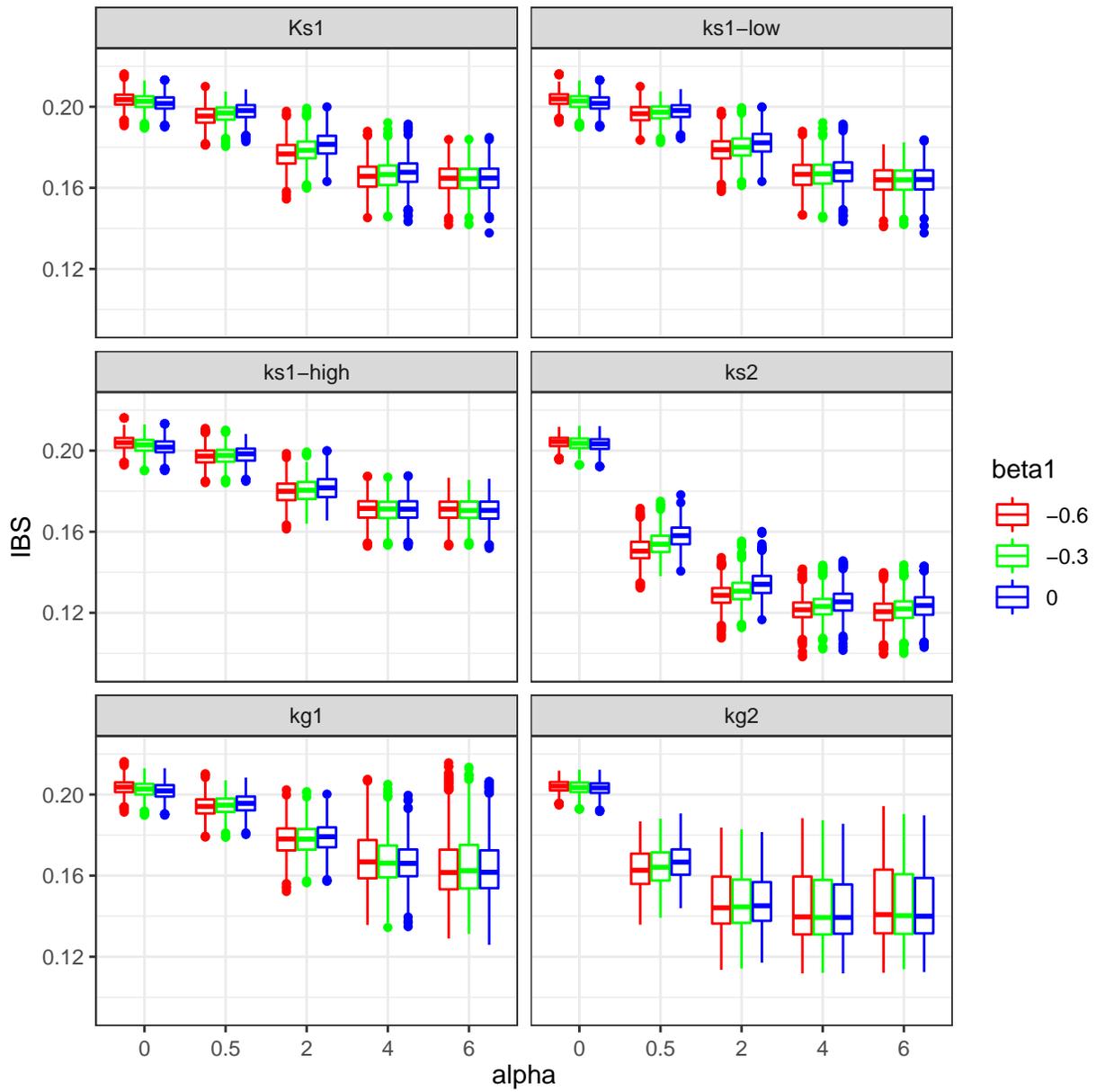

Figure S4: IBS distribution by alpha and beta_1

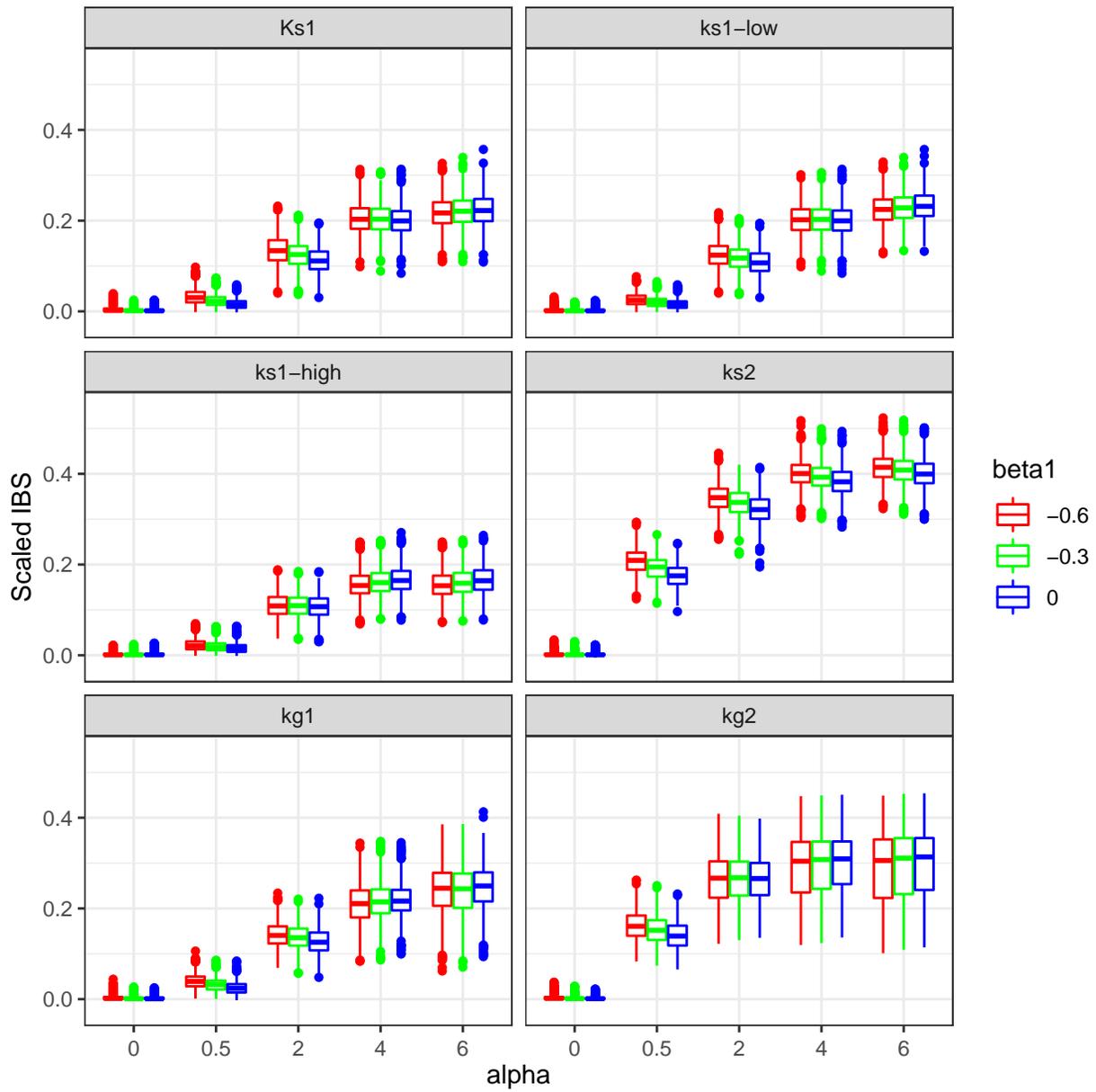

Figure S5: Scaled IBS distribution by alpha and beta_1

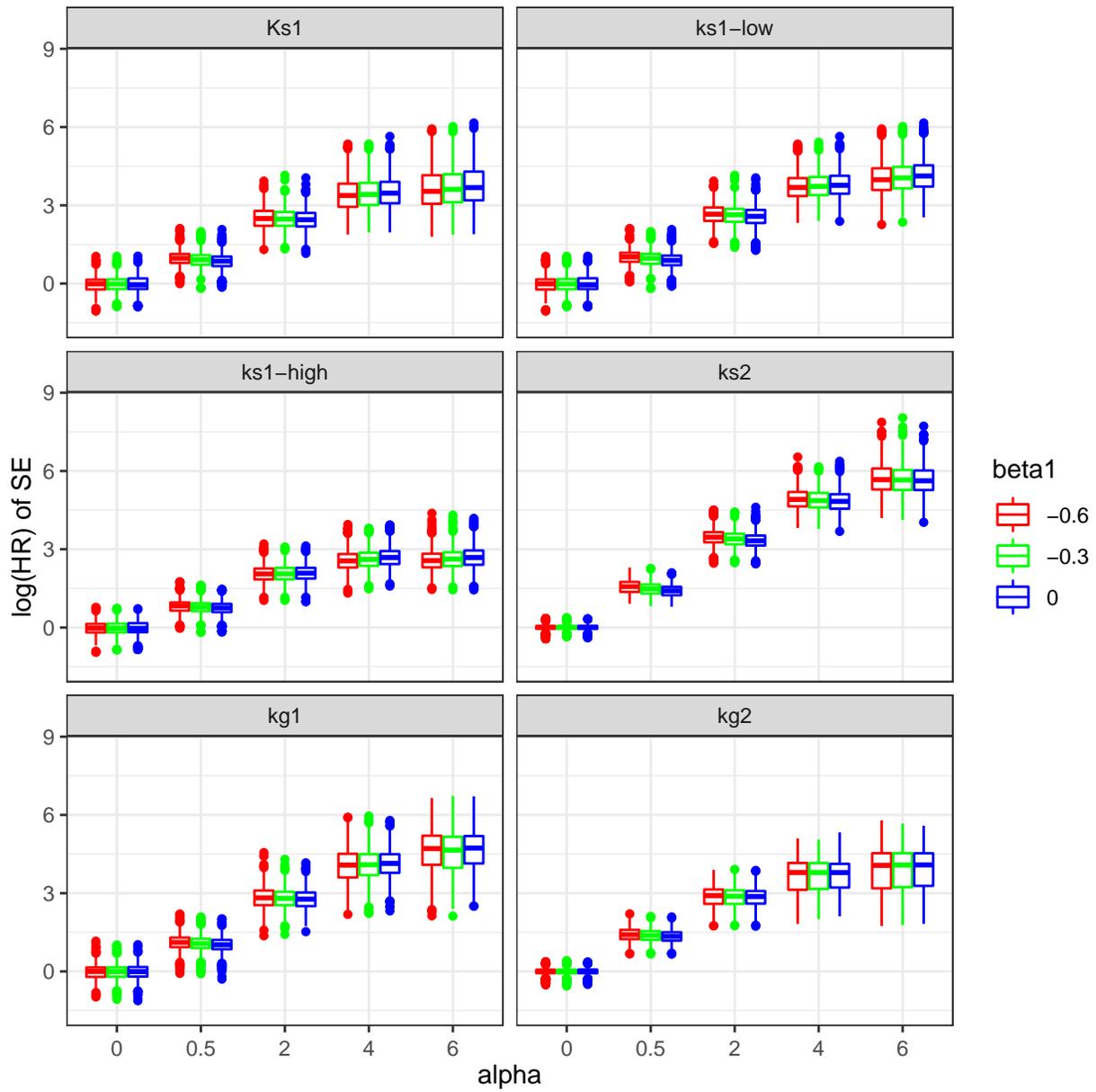

Figure S6: log(HR) distribution by alpha and beta_1

trial level association

**Grouped by unique values of alpha and beta1**

**sampling 3 replicates per Ks/Kg parameter**

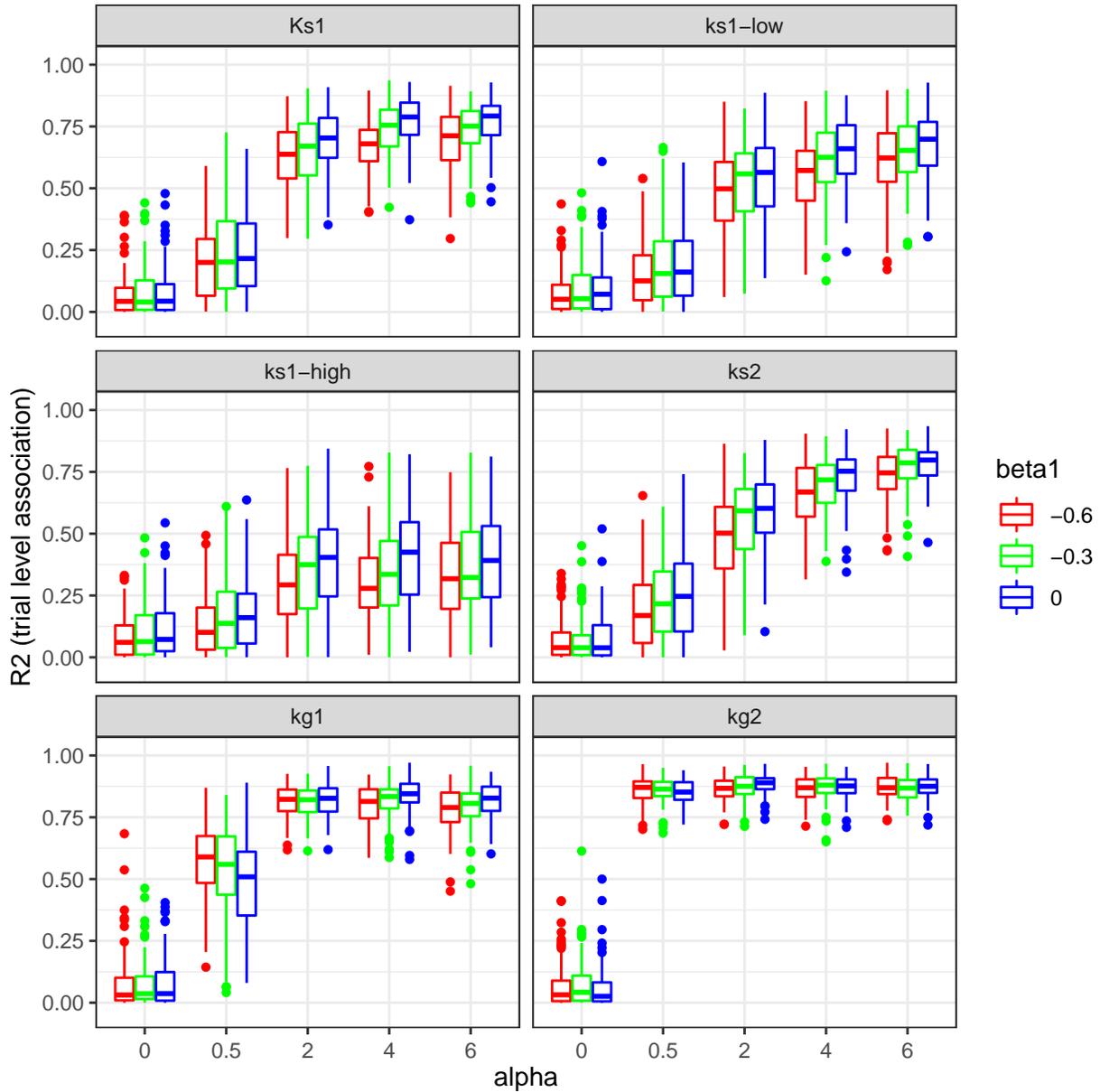

Figure S7: R2 by alpha and beta_1 from 15 studies

Each box summarizes R2 from 100 simulated set. Each set includes 15 studies. There are 3 duplicates for each of the 5 values of Ks/Kg parameter in the active arm.

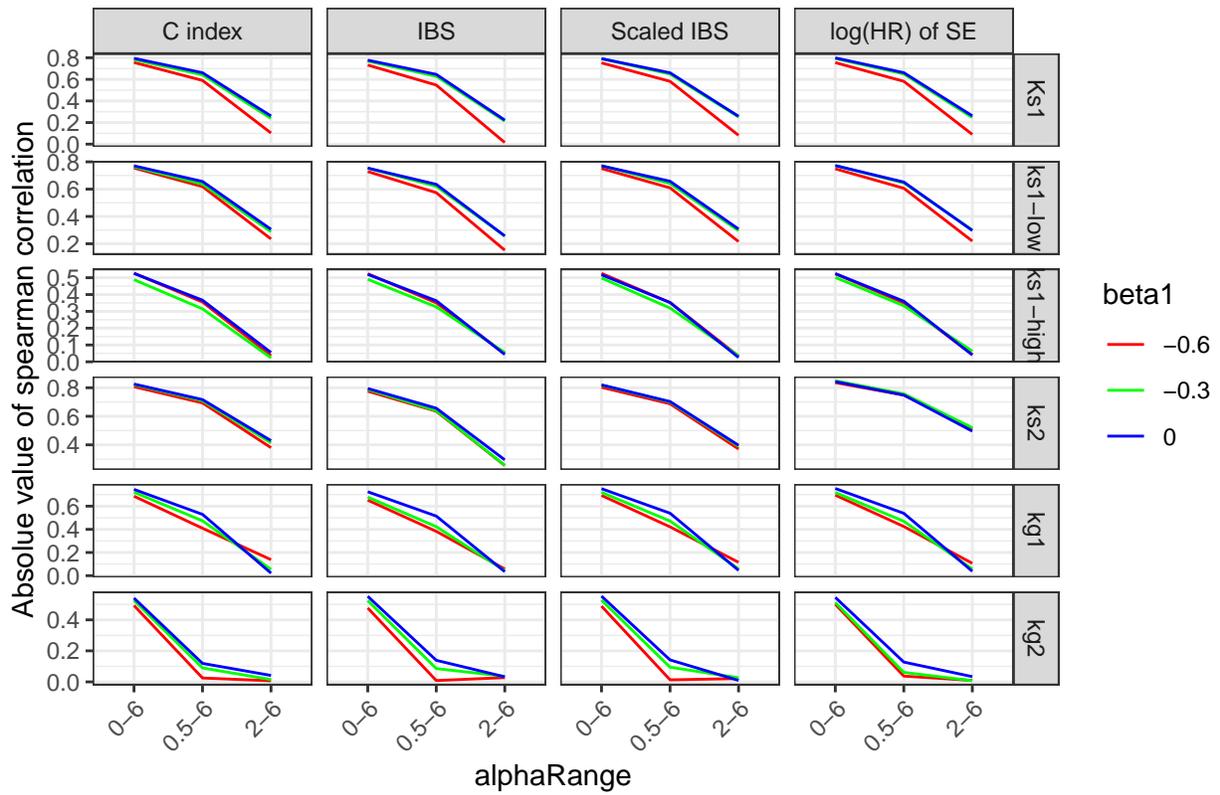

Figure S8: Between patient level and trial level association (dup 3)

**sampling 1 study per Ks/Kg parameter**

Figure S9: R2 by alpha and beta_1 from 5 studies

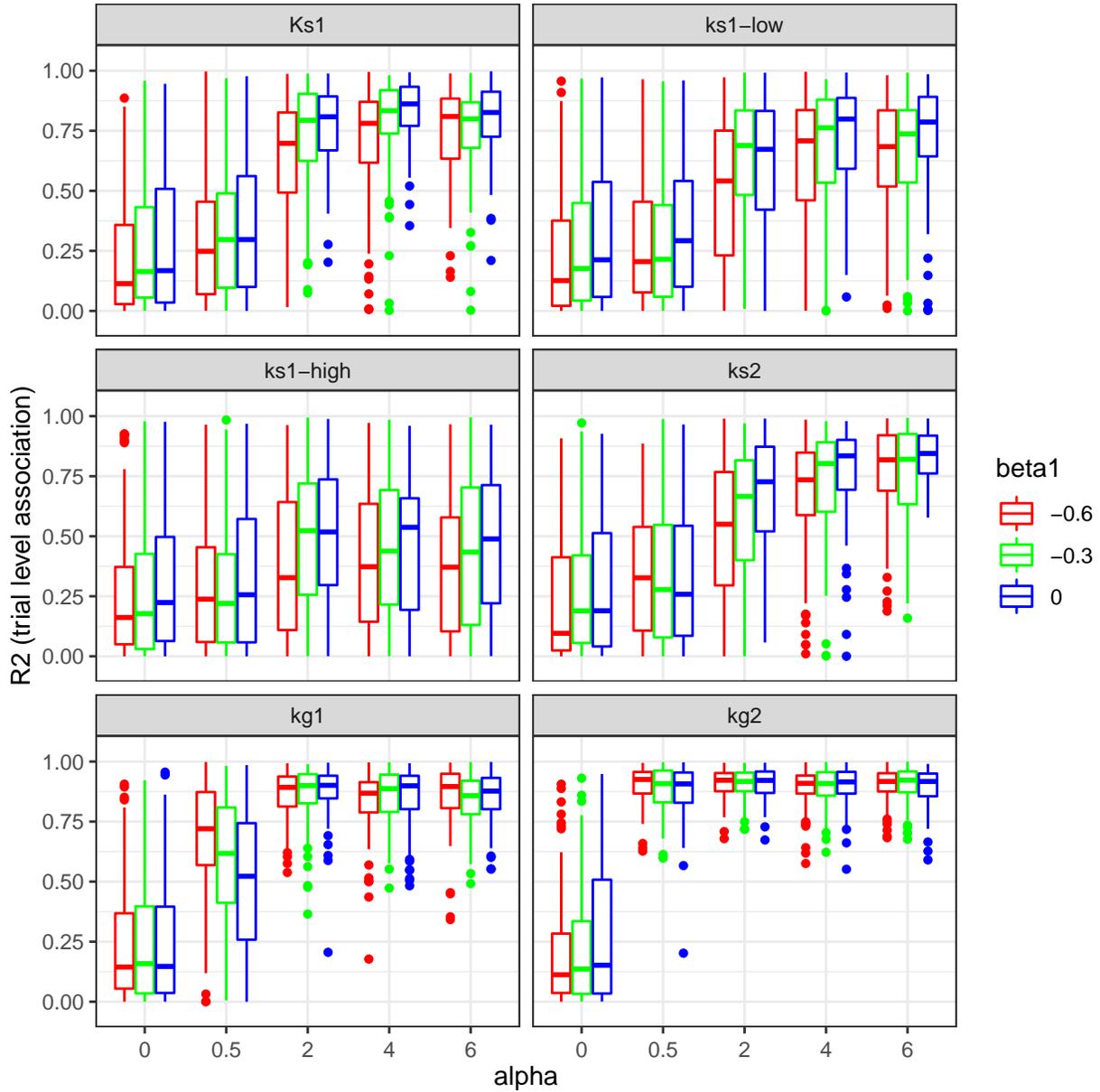

Each box summarizes R2 from 100 simulated set. Each set includes 5 studies. There are 1 duplicates for each of the 5 values of Ks/Kg parameter in the active arm.

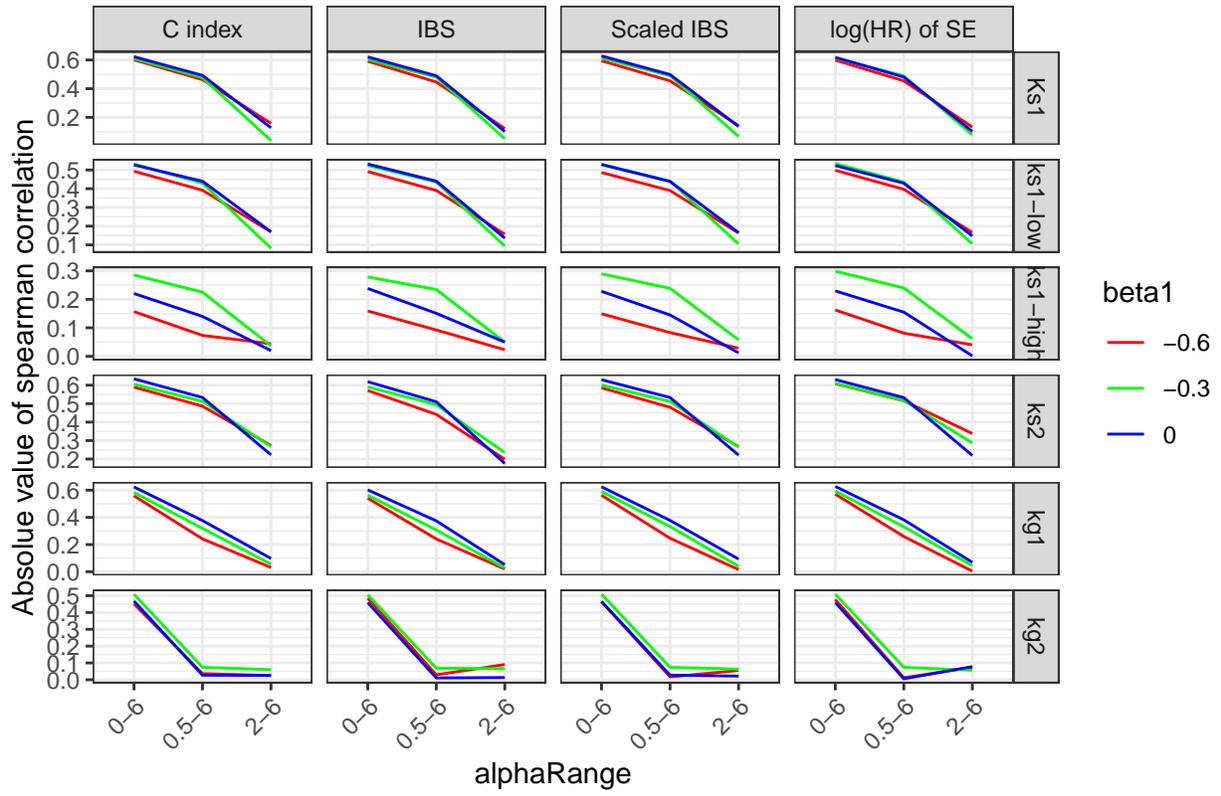

Figure S10: Between patient level and trial level association (dup 1)

**Grouped by alpha**

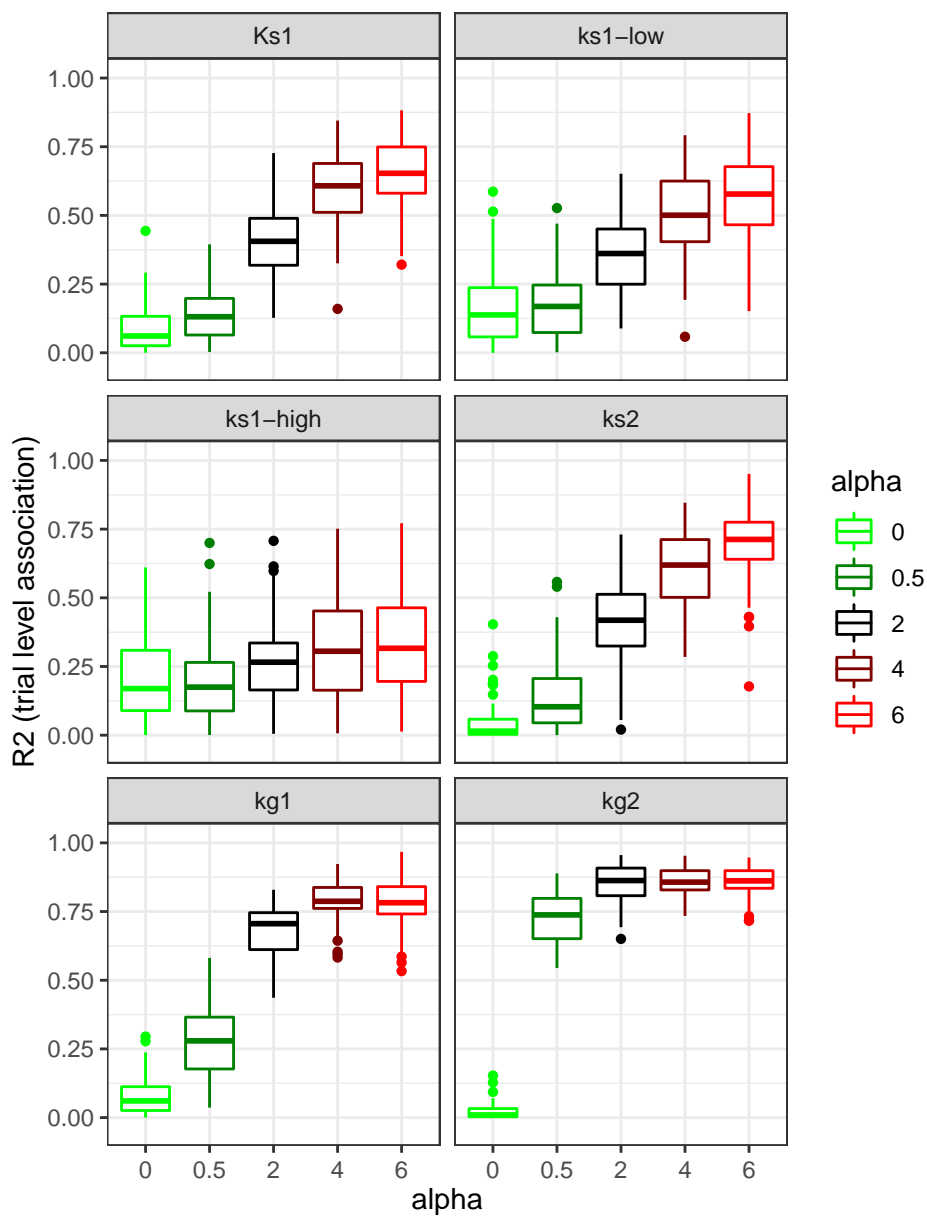

Figure S11: R2 by alpha from 15 studies

Each box summarizes R2 from 100 simulated set. Each set includes 15 studies. There are 1 duplicates for each of 3 beta_1 values and the each of 5 values of Ks/Kg parameter in the active arm.

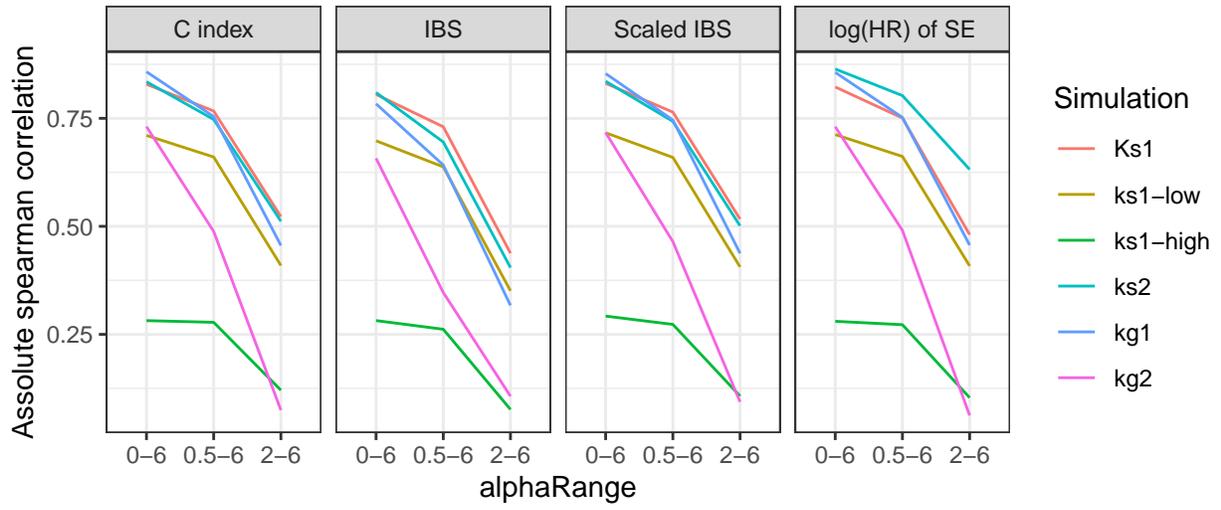

Figure S12: Between patient level and trial level association